\documentclass[
reprint,
twocolumn,
floatfix,
aip,
rsi,
superscriptaddress
]{revtex4-2}

\usepackage{graphicx}
\usepackage{amsmath}
\usepackage{amssymb}
\usepackage{hyperref}
\usepackage{booktabs}

\begin{document}

\title{PiMiX 2.02: Toward AI-Driven Data Fusion in Radiographic Imaging and Tomography}

\author{Zhehui Wang}
\email{Contact: zwang@lanl.gov}
\affiliation{\protect\footnotesize Los Alamos National Laboratory, Los Alamos, NM 87545, USA}

\author{Shanny Lin}
\affiliation{\protect\footnotesize Los Alamos National Laboratory, Los Alamos, NM 87545, USA}
\affiliation{\protect\footnotesize The University of Texas at Austin, Austin, TX 78758, USA}

\author{Nicholas Amano}
\affiliation{\protect\footnotesize University of Michigan, Ann Arbor, MI 48109, USA}

\author{Ramya Gurunathan}
\affiliation{\protect\footnotesize NVIDIA, Santa Clara, CA 95051, USA}

\author{Katie Liu}
\affiliation{\protect\footnotesize Brown University, Providence, RI 02912, USA}

\author{Nathan E. Peterson}
\affiliation{\protect\footnotesize Los Alamos National Laboratory, Los Alamos, NM 87545, USA}

\author{Michelle A. Espy}
\affiliation{\protect\footnotesize Los Alamos National Laboratory, Los Alamos, NM 87545, USA}

\author{Adam Thompson}
\affiliation{\protect\footnotesize NVIDIA, Santa Clara, CA 95051, USA}

\author{Amy J. Clarke}
\affiliation{\protect\footnotesize Los Alamos National Laboratory, Los Alamos, NM 87545, USA}

\author{Ray T. Chen}
\affiliation{\protect\footnotesize The University of Texas at Austin, Austin, TX 78758, USA}

\date{\today}

\begin{abstract}
PiMiX (Physics-informed Meta-instrument for eXperiments) was introduced for multi-instrument, multi-experiment, and simulation--experiment data fusion in radiographic imaging and tomography (RadIT). Here we present PiMiX~2.0 as an evolving artificial-intelligence (AI)-enhanced cyber-physical meta-instrument that integrates imaging sensors, near-sensor computing, multi-level data fusion, physics-informed inference, and human-supervised AI workflows across X-ray, neutron, and other modilities. PiMiX~2.0 treats measurement hardware, distributed computation, data analysis, and inference as harmonious elements of a meta-instrument. Demonstrated capabilities include multimodal Complementary Metal-Oxide-Semiconductor (CMOS) radiation imaging, simulation-assisted sub-pixel neutron localization, and edge-deployed optical-neural-network (ONN) inference; in a recent neutron-imaging demonstration, Graphics Processing Unit (GPU) and ONN implementations achieved greater than 96\% precision for neutron-event detection with sub-micron localization. Another key advance is human-in-the-loop agentic-AI co-analysis of X-ray and neutron images from inertial confinement fusion. Besides denoising, smoothing, Fourier transformation, and illumination correction, the workflow generates competing feature hypotheses, ranks contours using physics-informed evidence, estimates confidence, and presents multiple possibilities for human review. The same agentic architecture is adapted to different physics: X-ray analysis emphasizing dark nonuniform ring structures using deformable closed paths and multi-scale dark-band evidence, whereas neutron analysis targeting bright emission envelopes using background-subtracted fractional-emission levels, outward intensity gradients, enclosed emission, and cross-filter persistence. We further highlight automated comparison of an as-designed stereolithography model with an X-ray computed-tomography reconstruction of an additively manufactured metal lattice. Together, these examples illustrate a progression from AI-assistance in specific data processing tasks, such as coding, to autonomous multi-task multi-domain data fusion, such as scientific co-analysis. PiMiX~2.0 also provides a pathway toward PRISM, a RadIT scientific foundational model, and toward a greater integration of diagnostics, digital representations, inference, and experimental control.

\end{abstract}

\maketitle


\section{Introduction\label{sec:intro}}
Radiographic imaging and tomography (RadIT) use penetrating radiation, including X-rays, high-energy X-rays and gamma rays, neutrons, and energetic charged particles, to infer the internal materials structure, composition, or dynamics of otherwise inaccessible systems~\cite{Wang22:RadIT,Wang2023:URadIT}. Applications span high-energy-density (HED) physics and nuclear fusion, dynamic materials, nondestructive testing, additive manufacturing, nuclear and high-temperature plasma physics, and medicine. In many of these applications, the quantity of scientific interest is not observed directly. Instead, it must be inferred from measurements that are incomplete in space, time, energy, viewing angle, or particle species. RadIT is therefore fundamentally an inverse and data-fusion problem in addition to instrumentation and measurement.

In HED, nuclear fusion, and high-temperature plasma experiments, for example, rapidly evolving plasmas span large ranges of spatial, temporal, particle (including massless particles such as photons), and spectral scales. X-ray imaging and spectroscopy, neutron imaging and spectroscopy, gamma-ray measurements, and charged-particle radiography provide complementary rather than redundant views of the plasma state. X-ray measurements can constrain morphology, emissivity, opacity, temperature, and density; neutron measurements probe fusion production, burn dynamics, and fuel assembly; and charged-particle radiography can provide information on matter distributions and electromagnetic fields. Each diagnostic nevertheless represents only a partial projection of a complex physical system, with tradeoffs among spatial, temporal, and spectral resolution. Consequently, quantitative interpretation increasingly requires measurements from multiple diagnostics to be analyzed together, `co-analysis', with calibration and geometry information, material-response models, simulations, metadata, and prior physical knowledge.

Closely related problems occur throughout RadIT. In computed tomography (CT), incomplete-view, sparse-view, and limited signal-to-noise ratio measurements motivate model-based, iterative, and data-driven reconstruction approaches~\cite{Feldkamp1984,Jin2017,Adler2018,Arridge2019}. In dynamic imaging, sequences of measurements must be combined to reconstruct the evolution of objects or plasma states that cannot be completely sampled at a single instant. In manufacturing, prior design information can be compared or fused with X-ray CT measurements for dimensional metrology and defect characterization~\cite{Ziabari2023}. These examples motivate a broader concept of \emph{data fusion} (DF): information from multiple sensors, experiments, simulations, and prior models are combined to construct a more informative representation of the physical system than can be obtained from any individual measurement alone.

The original Physics-informed Meta-instrument for eXperiments (PiMiX) framework organized this concept into several kinds of DF, including multi-instrument data fusion (MIDF), multi-experiment data fusion (MXDF), and simulation--experiment data fusion (SXDF)~\cite{Wang2024:PiMiX}. MIDF combines complementary measurements of the same or closely related physical state; MXDF connects information across shots, campaigns, instruments, or facilities; and SXDF combines measurements with forward models or simulations for reconstruction, parameter inference, validation, and uncertainty assessment. In this formulation, the ``meta-instrument'' is not a single detector. Rather, it is the coordinated measurement, data processing and inference system formed by diagnostics hardware, their computational representations, and the models used to interpret their outputs. PiMiX~2.0 extends the conceptual DF in PiMiX to an AI-enhanced \emph{cyber-physical} framework and architecture for RadIT. Here, the term cyber-physical emphasizes that sensing, computation, data reduction, reconstruction, inference, and human interaction are treated as coupled elements of the measurement system rather than as a detector followed by an optional post-processing pipeline. 

The technological building blocks underlying the PiMiX 2.0 architecture have broad foundations. Near-sensor and low-latency machine learning, heterogeneous computing, physics-informed machine learning, multimodal data analysis, and AI-assisted scientific workflows are active research areas across high-energy physics, imaging, materials science, chemistry, and other disciplines~\cite{Aaij2019,Fahim2021,Pennicard2024,Karniadakis2021,Boiko2023,Bran2024,Szymanski2023}. Likewise, large multimodal and foundation-model concepts are increasingly being investigated for scientific and medical applications~\cite{Pai2024,Niu2025}. 

In this work, we describe the resulting PiMiX~2.0 architecture, key hardware components and initial demonstration of agentic AI-workflow for X-ray and neutron image analysis, driven by different physics considerations. The current system may be characterized as \emph{AI-enhanced}: demonstrated capabilities and prototype integrations coexist with elements of a longer-term \emph{AI-driven} architecture that remain under development. Sec.~\ref{sec:architecture} gives an overview of the PiMiX 2.0 architecture, new capabilities since the initial PiMiX~\cite{Wang2024:PiMiX}, and maturity status. Sec.~\ref{sec:hw} summarizes the multi-modal CMOS imaging hardware foundation, augmented by optical-neural-networks (ONN)-based edge computing. Sec.~\ref{sec:datafusion} discusses how machine learning and AI, agentic AI in particular, are transforming and accelerating DF and applications. Sec.~\ref{sec:app} highlights the results of agentic-AI-mediated X-ray and neutron data co-analysis with human supervision. In summary, these developments provide a pathway toward increasingly integrated AI-driven PiMiX and applications.
\section{PiMiX 2.0: Cyber-Physical Integration Framework \label{sec:architecture}}
\begin{figure*}[htbp]
\centering
\includegraphics[width=0.9\textwidth]{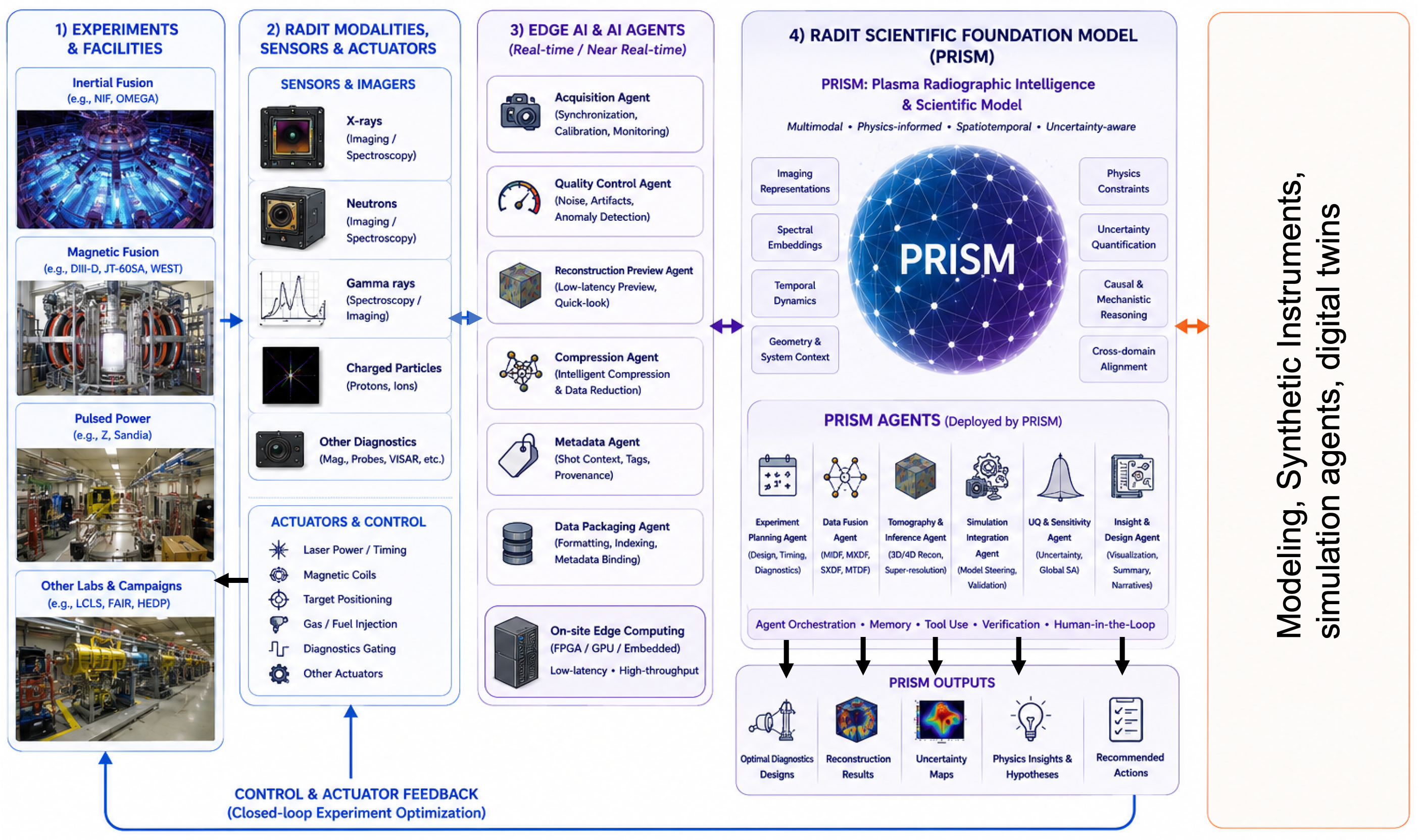}
\caption{Reference architecture of PiMiX~2.0 as an AI-enhanced cyber-physical meta-instrument. Measurements and associated experimental metadata enter through RadIT sensors and may be processed using detector-adjacent computing before transfer to workstation, GPU, or HPC resources. Higher-level processing supports reconstruction, MIDF, MXDF, SXDF, simulation, and physics-informed inference. Human-supervised AI tools assist analysis and workflow orchestration. PRISM denotes a developing multimodal RadIT scientific-model layer intended to connect measurements, simulations, experimental context, and scientific knowledge. The return path from inference to experimental decision and control is a longer-term architectural target that will extend the current work.}
\label{fig:PiMiX2}
\end{figure*}

Leveraging rapid advances in AI, including large language models (LLMs), agentic and other AI systems, and scientific machine learning, PiMiX~2.0 extends the original PiMiX from a DF concept toward an AI-enhanced cyber-physical integration architecture for RadIT, as illustrated in Fig.~\ref{fig:PiMiX2}. The central goal is to treat the physical measurement system, computational infrastructure, and scientific inference workflow as coupled harmonious elements of the meta-instrument, with the objective of accelerating data flow, information extraction, and experimental decision making. 
This hierarchical architecture is intended to reduce unnecessary data movement and analysis latency while allowing computational effort to be matched to the complexity of each task. 

\subsection{New capabilities in PiMiX 2.0}

PiMiX~2.0 retains the DF foundation in our earlier work~\cite{Wang2024:PiMiX} and extends the meta-instrument concept further in two directions. The first extension moves selected computation closer to the measurement~\cite{LZSM:2025}. High-rate radiation imaging can approach or exceed practical data-transfer and archival constraints, motivating extraction of scientifically useful information at or near the detector. PiMiX~2.0 therefore incorporates a hierarchical computing model in which operations such as event detection, localization, feature extraction, quality assessment, or data reduction may be performed using detector-adjacent electronic or optical computing, while more computationally intensive reconstruction, cross-diagnostic DF, simulation, and inference are performed using progressively larger computing resources. Recent CMOS neutron-imaging and ONN results provide experimental foundations, including quantitative event-detection and localization performance~\cite{LBBC:2023,LZSM:2025}. Full real-time integration of the detector, optical inference hardware, higher-level DF, and experimental feedback remains under development.

The second extension introduces human-supervised AI-assisted scientific workflows. LLMs and other AI tools can now routinely assist with code generation, execution of established analysis routines, data transformation, visualization, and coordination of complex wokflows in data processing. Within PiMiX~2.0, these capabilities complement rather than replace physics models, uncertainty analysis, and expert validation. New results are presented in Sec.~\ref{sec:app}. 

Together, these extensions make the computational infrastructure and agentic-AI workflows an integral part of the RadIT measurement chain rather than a disposable post-processing stage. At the physical measurement layer, PiMiX~2.0 is intended to interface with RadIT diagnostics based on X rays, neutrons, gamma rays, and energetic charged particles. The architecture is detector-agnostic in principle, although the present work is motivated particularly by CMOS and active-pixel imaging systems~\cite{WBHM:2016,PFWG:2016,WABD:2021,Wang2023:URadIT}. 

The detector-adjacent layer provides low-latency processing and data reduction before data are transferred to higher-level computing resources. Edge computing itself is not unique to PiMiX; related approaches are established in high-energy-physics triggering, FPGA- and ASIC-based machine-learning accelerators, and high-rate photon-science data reduction~\cite{Aaij2019,Fahim2021,Pennicard2024}. Within PiMiX~2.0, the purpose of this layer is to generate scientifically useful intermediate information, for example,  detected events, estimated interaction positions, regions of interest, compressed representations, or data-quality indicators, while retaining sufficient raw information and provenance for subsequent ingestion. ONN is also attractive in terms of low power consumption and radiation hardness.

Higher-level computing resources support operations that require greater computational capacity or information from multiple measurements. These may include tomographic reconstruction, image registration, feature extraction, MIDF and MXDF, synthetic diagnostics, uncertainty estimation, SXDF, and physics simulation. PiMiX~2.0 does not prescribe a single reconstruction algorithm, solver, machine-learning architecture, or computing platform. Instead, it emphasizes interfaces among measurements, geometry, models, algorithms, and derived data products, together with sufficient provenance to establish how a scientific result was obtained. This requirement becomes especially important when learned models or AI-generated code are introduced into the analysis workflow and is consistent with broader principles of reproducible scientific data and computation~\cite{Wilkinson2016}.

Beyond PiMiX 2.0 demonstrated capabilities, PRISM (\emph{PiMiX Radiographic Intelligence and Scientific Model}) is envisioned as a multimodal model layer within the staged architecture, or the `AI Master' (AIM), see Sec.~\ref{sec:datafusion} for additional discussions. Its intended role is to relate RadIT measurements to experimental metadata, geometry, simulations, synthetic data, and relevant scientific knowledge, providing shared representations that can support DF and AI-assisted scientific workflows. In the present work, PRISM denotes a developing architectural capability rather than a completed production-scale foundation model. The architecture also includes a prospective return path from scientific inference to experimental decision and actuation. Reconstructed observables and simulation--experiment comparisons could, for example, inform diagnostic configuration, timing, experimental design, or other controllable parameters. Digital twins and synthetic diagnostics provide one mechanism for evaluating candidate configurations before experimental changes are implemented. No end-to-end closed-loop facility-control experiment is reported here; diagnostic-to-control integration therefore remains an architectural target. Near-term PiMiX~2.0 development emphasizes human-guided operation, with domain experts retaining responsibility for scientific validation and experimental decisions.

\subsection{Maturity and implementation status}

Because the components of PiMiX~2.0 are at different stages of development, we distinguish three levels of maturity, as summarized in Table~\ref{tab:status}. \emph{Demonstrated} components have been evaluated experimentally or computationally; \emph{prototype} capabilities have yielded new results in the workflows; and \emph{targets} identify capabilities under development that have not yet been demonstrated as an integrated PiMiX~2.0 system.

\begin{table}[!ht]
\centering
\caption{Maturity of key PiMiX~2.0 elements.}
\label{tab:status}
\begin{tabular}{@{}p{0.64\columnwidth}p{0.26\columnwidth}@{}}
\toprule
\textbf{PiMiX~2.0 element} & \textbf{Status} \\
\midrule
Multi-modal CMOS imaging                              & Demonstrated \\
NN-assisted neutron localization          & Demonstrated \\
ONN edge inference                        & Demonstrated \\
AI-assisted code generation   & Prototype \\
Agentic AI workflow   & Prototype \\
(\textit{Human-in-the-loop}) & \\
PRISM                                     & Target \\
Digital-twin \& closed-loop control       & Target \\
\bottomrule
\end{tabular}
\end{table}
\section{Imaging and edge-computing hardware \label{sec:hw}}
The hardware foundations of PiMiX for RadIT are CMOS imaging sensors and edge computing capabilities. One recent key advancement in PiMiX 2.0 is the expansion to integrate ONN hardware.

\paragraph{Multimodal CMOS imaging sensors} CMOS and related active-pixel sensors provide a useful hardware basis for PiMiX because they have been applied to multiple radiation modalities, including X-ray, neutron, and charged-particle detection~\cite{WBHM:2016,PFWG:2016,WABD:2021,LBBC:2023,LZSM:2025}. Rather than attempting to catalog every camera tested by the collaboration, Table~\ref{tab:camera_comparison} lists representative systems that illustrate the range from high-pixel-count continuous imaging to ultrafast burst acquisition.

\begin{table*}[htbp]
\centering
\caption{Representative CMOS-based camera systems used in high-speed or radiation-imaging studies. Values are retained from the original manuscript.}
\label{tab:camera_comparison}
\begin{tabular}{lccccccc}
\toprule
Camera & Pitch ($\mu$m) & Rows & Columns & Bits & Frame rate (fps) & Burst/CW & Example RadIT mode \\
\midrule
Nikon D800 & 4.88 & 4912 & 7360 & 14 & $\sim 4$ & CW & $\beta$ (Sr-90) \\
Shimadzu HPV-X2 & 32 & 400 & 250 & 10 & $\leq 10^{7}$ & 256 frames & X ray (20--30 keV) \\
PCO DiCam Pro C1 & 12 & 1008 & 1008 & 12 & $\leq 10^{7}$ & 2 frames & X ray ($\leq 60$ keV) \\
\bottomrule
\end{tabular}
\end{table*}

A recurring systems challenge is the mismatch between detector data generation and sustainable data movement or processing. Similar constraints motivate real-time reduction in photon science and trigger processing in high-energy physics~\cite{Pennicard2024,Aaij2019}. In PiMiX, this motivates but does not mandate detector-adjacent processing: the appropriate solution may be buffering, conventional compression, region-of-interest selection, GPU/FPGA inference, optical inference, or a combination determined by the experiment.

\paragraph{Edge computing and optical neural networks}
Optical neural networks (ONNs) are one candidate edge-computing technology within the PiMiX architecture; they are not assumed to be the unique or universally optimal solution. The broader field includes electronic GPUs, FPGAs, and ASICs with mature scientific use cases~\cite{Aaij2019,Fahim2021}. The current PiMiX emphasis for exploring photonic accelerators is to evaluate whether their matrix--vector processing characteristics and potential radiation tolerance can be useful for edge RadIT data reduction~\cite{NZFG:2025,LZSM:2025}.

\begin{figure}[htbp]
    \centering
    \includegraphics[width=0.45\textwidth]{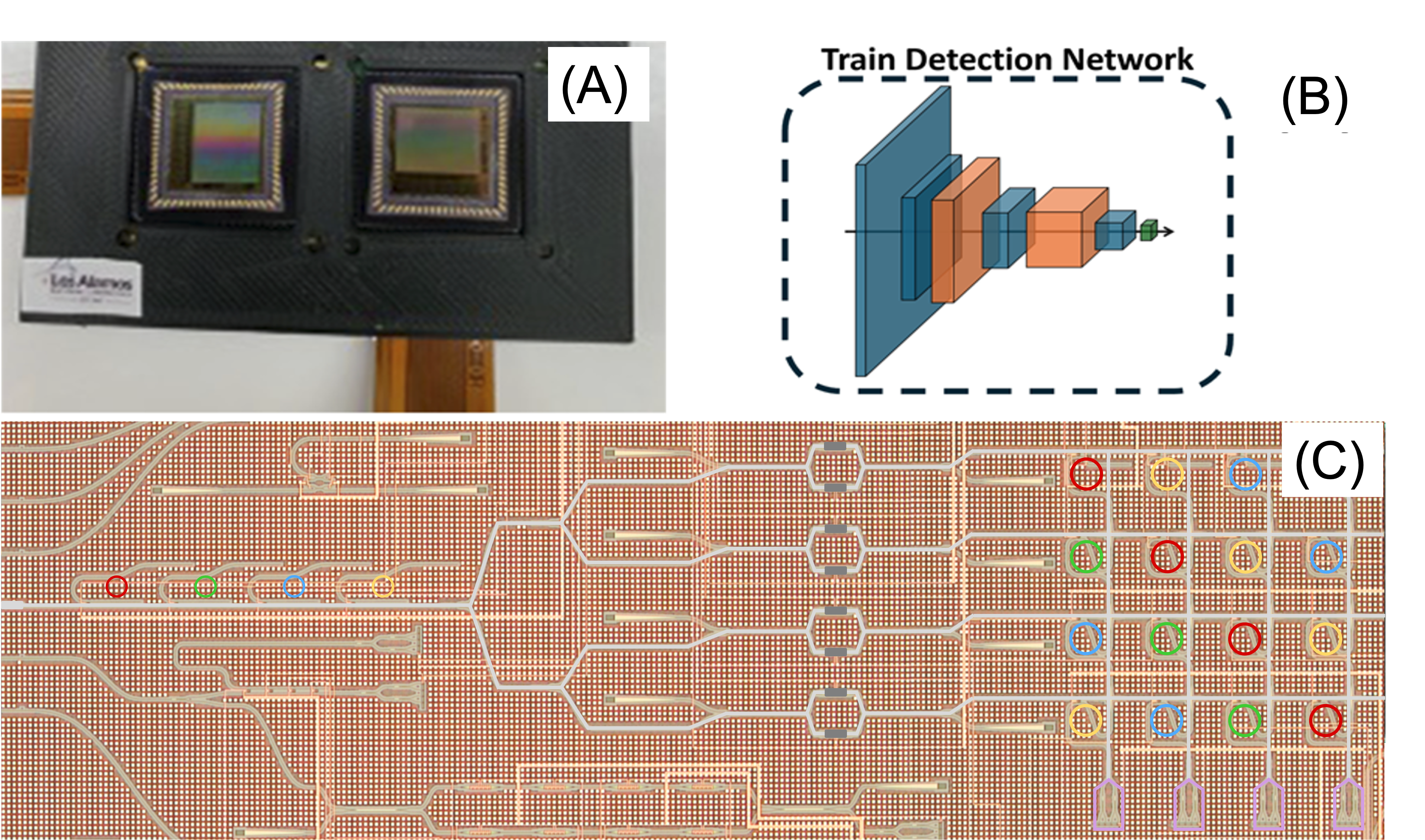}
    \caption{CMOS sensing and optical edge-computing elements integration in the PiMiX program. (a) CMOS image sensors used in a tiling configuration for wide-field X-ray imaging~\cite{WABD:2021}. (b) Neural-network workflow for detector-adjacent processing~\cite{LZSM:2025}. (c) A prototype optical-neural-network (ONN) circuit, Courtesy of Dr. Shupeng Ning. More details about ONN may be found in Ref.~\cite{NZFG:2025} and cited work within.}
    \label{fig:CMOS1}
\end{figure}

\section{AI-enhanced data fusion \label{sec:datafusion}}

AI algorithms and agentic AI assistance, as a cross-cutting theme, can enhance DF at various stages of
data collection, processing, modeling, and interpretation. At the data collection and preprocessing stage, neural network methods can assist calibration,
registration, denoising, data compression, segmentation, feature extraction, or high-dimension reconstructions from low-dimension projections.
At the modeling stage, machine-learning models, scientific foundation models,
surrogates, or simulation-derived representations can participate in the
experimental data post-processing and understanding. The complexity of the growing AI tasks and varieties motivates orchestrated AI activities or `\textit{task fusion}', where one or more AI agents, under the supervision of a human or `AI Master' (AIM), can select and
invoke analysis tools, explore hypotheses or parameter choices,
compare intermediate results, maintain workflow state and provenance, and
request human intervention when, for example, predefined criteria are not satisfied. DF and AIM development can further leverage scientific and multimodal
foundation models~\cite{Pai2024,Niu2025}, with RadIT-specific implementation and constraints, including explicit detector geometry and
response, heterogeneous and often sparse labels, radiation-specific uncertainty, provenance, and facility
dependencies.  

\subsection{Synthetic--experimental DF for neutron measurements}

Recent neutron-imaging work  provides an earlier example of AI-enhanced
SXDF at the detector level~\cite{LBBC:2023,LZSM:2025}. Experiment-like CMOS neutron-hit images with known
interaction coordinates were generated using detector simulation and used to
train localization networks, while experimental measurements were used to
evaluate realism and practical applicability. This combination addresses an
important limitation of experimental supervision: the precise neutron
interaction coordinate required as ground truth for sub-pixel localization is
generally unavailable directly from measured detector images.

The follow-on edge-computing study extended this approach from localization alone
to a detector-plus-localizer workflow and evaluated electronic GPU and OSENN
implementations~\cite{LZSM:2025}, and achieved greater than 96\% precision, greater than 98\% recall,
greater than 97\% F1 score for neutron-event detection, together with
sub-micron localization. Repeated stochastic forward passes were additionally used to estimate
distributions of predicted interaction coordinates~\cite{LZSM:2025}, providing
a specific mechanism for associating uncertainty with the learned inference.

This SXDF example is helpful for distinguishing AI-enhanced task excution
from agentic AI workflows. The neural network performs a well-defined learned mapping
within a predetermined workflow; it does not decide which scientific tool to
run next, revise the analysis objective, launch an alternative simulation, or
evaluate competing physical hypotheses. It therefore represents an
AI-enhanced component within SXDF, but not an agentic workflow. The progression
toward PiMiX~2.0 and beyond is to preserve such quantitatively validated components while
making their composition, comparison, and iterative use increasingly adaptive
and traceable.

\subsection{AI agents and human supervision}

The ecosystem of AI agents for both technical and mundane yet time-consuming tasks provides a broader context for
this direction. An AI agent here refers not simply to an
LLM responding to a prompt, but to algorithms that can pursue a user-specified scientific
goal that requires integration of multiple steps, maintaining relevant states, invoking external tools or
models, evaluating intermediate observations, and designing subsequent actions
on those observations. Scientific-agent systems are now increasingly used in
chemistry, autonomous synthesis, hypothesis generation, computational
experimentation, multi-agent scientific reasoning, and end-to-end research
workflows~\cite{Boiko2023Coscientist,Bran2024ChemCrow,
Szymanski2023ALab,Lu2024AIScientist,Ghafarollahi2024SciAgents,
Schmidgall2025AgentLab,Gottweis2025CoScientist}. A transition from AI for specific dataset
inference or generation toward ``agentic science'' has been recognized~\cite{Xin2025AgenticScience}.

A specific example is URSA (Universal Research and
Scientific Agent), developed at Los Alamos~\cite{Grosskopf2025URSA}. URSA uses modular, composable agents for
activities including planning, hypothesis generation, research, code execution,
and interaction with scientific computational tools. Its demonstrated
architecture also includes coupling to advanced physics simulations, including
radiation-hydrodynamics calculations relevant to inertial confinement fusion (ICF). This illustrates a
general direction that is highly compatible with PiMiX: an agent need not
replace trusted scientific software, but can instead coordinate validated
simulation and analysis tools while retaining domain scientists as supervisors.

There is nevertheless significant room for
growth in unrestricted autonomous discovery, in comparison to highly specific and bounded scientific tasks. Current
systems have demonstrated important capabilities in literature-guided
reasoning, tool selection, code generation, computational experiments,
materials and chemistry workflows, and hypothesis generation, but reliability,
verification, reproducibility, uncertainty propagation, robustness to
tool/model errors, and accountability remain active research problems
\cite{Xin2025AgenticScience,Grosskopf2025URSA}. In experimental science these
issues become more consequential because an incorrect intermediate result can
propagate from software into physical decisions. Human supervision should
therefore be treated as part of the system architecture rather than merely as
a fallback after an agent fails.

Therefore, the most credible near-term role for agentic AI in PiMiX 2.0 and upgrades  is not
to replace the scientist, nor to provide an unrestricted path from an LLM to
facility control. It is to reduce the human effort required to coordinate
heterogeneous data and trusted scientific tools while making intermediate
decisions, assumptions, and alternatives more visible. Progress can then be
evaluated quantitatively through task success, analysis latency, number and
type of human interventions, reproducibility, error rate, agreement with
reference analyses, uncertainty calibration, and ultimately the accuracy and
precision of the inferred physical quantities. Such measurements provide a
staged path from AI-assisted computation toward increasingly capable
human--AI scientific collaboration.

Sec.~\ref{sec:app} below provide results for agentic-AI workflow
in PiMiX 2.0. They illustrate an iterative interaction
between human scientific inputs and AI-assisted analysis rather than a
fully autonomous scientific process. The detailed Gated X-ray Detector (GXD) and Neutron Imaging System (NIS) examples are
discussed next. 

\section{PiMiX 2.0 Demonstrations \& Prototypes \label{sec:app}}

This section highlights three representative PiMiX~2.0 prototype workflows. We first present a human-supervised agentic-AI analysis of GXD images from the National Ignition Facility (NIF), where the workflow evolves from conventional filtering and contour extraction to adaptive, physics-informed identification of weak shell-like features. We then extend the same agentic framework to co-analysis of NIS data, with the feature model reformulated to identify bright neutron-emission envelopes rather than dark X-ray shell boundaries. Finally, we demonstrate automated comparison between three-dimensional as-designed models and X-ray computed-tomography reconstructions for validation of additively manufactured structures. Together, these examples illustrate how PiMiX~2.0 integrates AI-assisted code generation, diagnostic-specific image analysis, human-in-the-loop refinement, and automated model-to-measurement comparison within a common DF framework.

\subsection{Agentic AI for X-ray image segmentation and analysis}

The analysis of spatial structures in X-ray and neutron imaging is challenging when the signal of interest is embedded in strongly nonuniform illumination, stochastic high-intensity fluctuations, detector noise, and partially resolved structures. In the initial PiMiX workflow, image features were characterized using conventional image-processing methods, including Fourier-domain analysis, illumination flattening, spatial filtering, and iso-intensity contours. These methods are deterministic, computationally efficient, and physically interpretable; however, their effectiveness depends strongly on local signal-to-noise ratio and on the assumption that the physical feature of interest can be represented by an approximately constant intensity level. The PiMiX 2.0 revision therefore provided an opportunity to explore an AI-agentic workflow in which the analysis procedure itself was iteratively evaluated and modified in response to observed failure modes.

\begin{figure*}[htbp]
    \centering
    \includegraphics[width=0.85\textwidth]{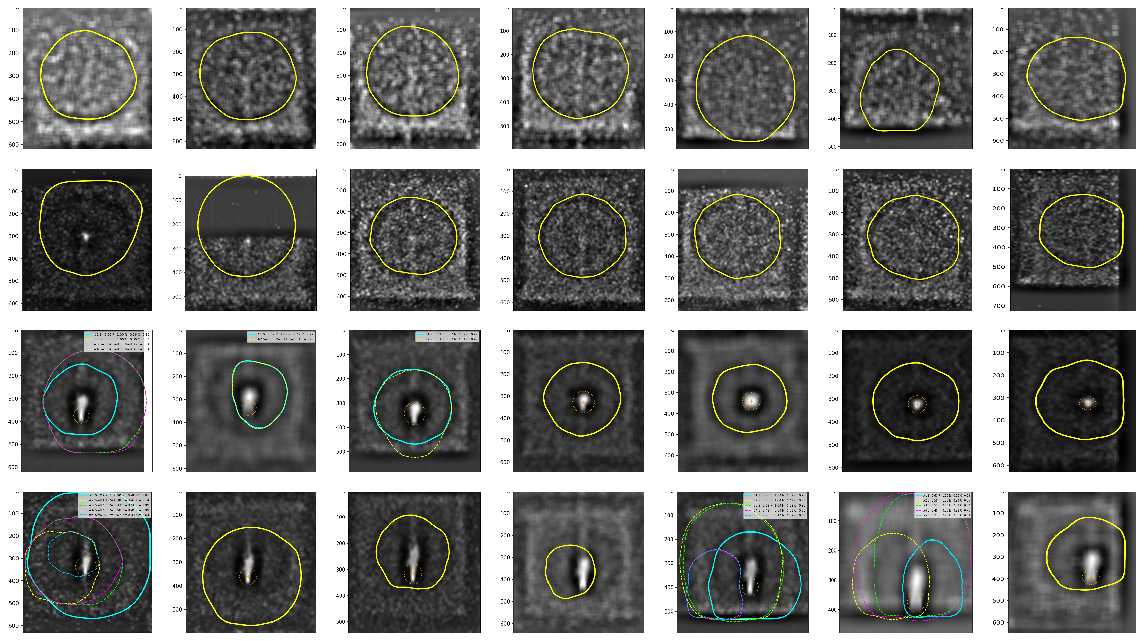}
    \caption{Human-in-the-loop agentic-AI contour analysis of a GXD image dataset from NIF. Each panel shows the inferred shell contour (or a few possible contours) on the top of AI-assisted preprocessed raw data. The workflow combines automated preprocessing, multi-scale feature enhancement, deformable closed-path search, and candidate contour ranking, with human supervision used to assess and refine difficult cases involving low signal-to-noise, bright central emission, or competing ring-like features.}
    \label{fig:Contour}
\end{figure*}

The starting workflow first segemented each raw data files (.h5 format) into user-guided spatial segments and applied local illumination flattening to reduce large-scale gradients. Several low-pass filters were subsequently implemented and evaluated, including a hard Fourier-mode cutoff, a smooth Butterworth Fourier filter, spatial box averaging, Gaussian smoothing, and median filtering. These methods substantially suppressed isolated high-intensity fluctuations and revealed large-scale structures that were difficult to recognize in the original image. Contours were then calculated from the filtered raw and flattened images. Although this approach improved visualization, it exposed a fundamental limitation of conventional iso-intensity contouring: a physical boundary can remain visually recognizable to a human even when its intensity changes substantially around its circumference. Consequently, no single intensity threshold can reliably recover a broken, nonuniform dark ring from such images.

This failure motivated a transition from pixel-intensity processing to physics-informed feature-aware inference. The first intelligent feature finder used locally normalized edge information and robust geometric fitting to identify disconnected image fragments that could belong to a common ring feature. A RANSAC-based circular model demonstrated that spatially disconnected features could be combined into a global structure, but also revealed the limitations of imposing an incorrect geometric prior: experimentally observed rings were frequently distorted and could not be accurately represented by perfect circles. The agentic-AI development loop therefore modified the underlying assumption rather than merely adjusting filter parameters.

The next stage replaced the rigid circular model with a deformable closed-path representation. Instead of fitting a predefined circle or ellipse, the image was transformed into a local dark-ridge likelihood map and represented in polar coordinates. A globally optimized closed path, r($\theta$) in two-dimensional polar coordinates, was then determined using dynamic programming that balanced image evidence against local shape smoothness. This allowed the inferred boundary to deform continuously with angle and to bridge regions in which the physical ring was weak or partially absent. The only remaining geometric constraint was approximate closed-loop topology, rather than circularity. This change encoded the scientifically useful prior, namely, the closed-loop structure, without prescribing a detailed shape.

An additional challenge was that successful detection still depended on manually choosing the filtering scale, allowed radius range, and shape flexibility. The subsequent agentic steps therefore shifted from manual parameter selection to autonomous hypothesis testing. Multiple smoothing scales, radial search intervals, and deformation constraints were evaluated automatically. Rather than selecting the first plausible solution, geometrically similar solutions were grouped and their persistence across different preprocessing choices were evaluated. The same recovered geometry was then displayed on both the raw and flattened data so that preprocessing influenced the detection process without obscuring its relationship to the original measurement. This procedure reproduced an important element of human image interpretation: confidence increases when the same physical structure remains identifiable for different data visualization techniques.

Follow-on agentic workflow iterations further addressed images containing several plausible closed structures. Instead of generating a single result, the algorithm retained multiple competing hypotheses and ranked them using physically motivated quantities. In particular, a candidate was rewarded when its centerline followed a finite-width dark valley with brighter image intensity on both its inward and outward sides. Candidate quality also incorporated angular coverage, agreement between raw and flattened data, smoothness, distance from artificial image boundaries, and persistence across filtering scales. The best hypothesis was explicitly compared with the runner-up. Thus, when several rings were nearly equally plausible, the software could report an ambiguous or unreliable detection rather than presenting an arbitrary contour as a confident result. Candidate visualization additionally preserved human-in-the-loop inspection for difficult cases.

Later on, further workflow tuning was implemented for images containing an intense localized source (X-ray self-emission from the highly imploded plasma) surrounded by a comparatively weak dark ring. Such bright structures created high-contrast gradients that could dominate otherwise valid ring evidence. The results as summarized in Fig.~\ref{fig:Contour}, thus introduced a core-aware, multi-scale dark-band representation. Compact bright regions are first detected from their local contrast, after which a tapered exclusion region suppresses the immediate high-contrast halo during ring searching without imposing the bright-source position as the geometric center of the ring. Dark-band evidence is evaluated over several spatial scales and normalized by local image variance, effectively providing a local signal-to-noise measure rather than an absolute intensity measure. Evidence is then coherently integrated around a candidate closed path. A weak but spatially coherent ring can therefore accumulate substantial support even when no individual portion of the ring has high contrast.

The final ranking among multiple closed loops  therefore considers dark-band signal-to-noise, two-sided valley contrast, raw/flattened consistency, cross-scale geometric persistence, boundary penalties, candidate competition, and, when applicable, physically plausible core enclosure. The algorithm also retains the ability to reject an image for insufficient or ambiguous evidence.

In summary, the agentic iterations for the GXD dataset consisted of repeated cycles of observation, diagnosis, hypothesis generation, algorithm modification, and human expert evaluation. For example, random intensity spikes motivated spatial filtering; spatially varying ring intensity exposed the limitations of iso-contours; noncircular structures invalidated rigid circle fitting; sensitivity to manually selected parameters motivated autonomous parameter exploration; multiple plausible rings motivated competing-hypothesis ranking; and bright central sources motivated core-aware dark-band inference. Human evaluation of intermediate results provided feedback to the next computational strategy while the resulting algorithms remained transparent and physically interpretable.

On the evaluation data set of 28 image segments, as shown in Fig.~\ref{fig:Contour}, the number of satisfactory results has steadily increased from $\sim$ 10, to approximately 16, and now more than 20. The remaining difficult cases are generally associated with very low signal-to-noise closed loops, strong competing structures, or ambiguous multiple-loop configurations. For these cases, candidate visualization and manual deformable-ring controls are retained so that autonomous inference augments rather than replaces scientific judgment.

\subsection{Co-analysis of NIS images}

\begin{figure*}[!tbp]
    \centering
    \includegraphics[width=0.85\textwidth]{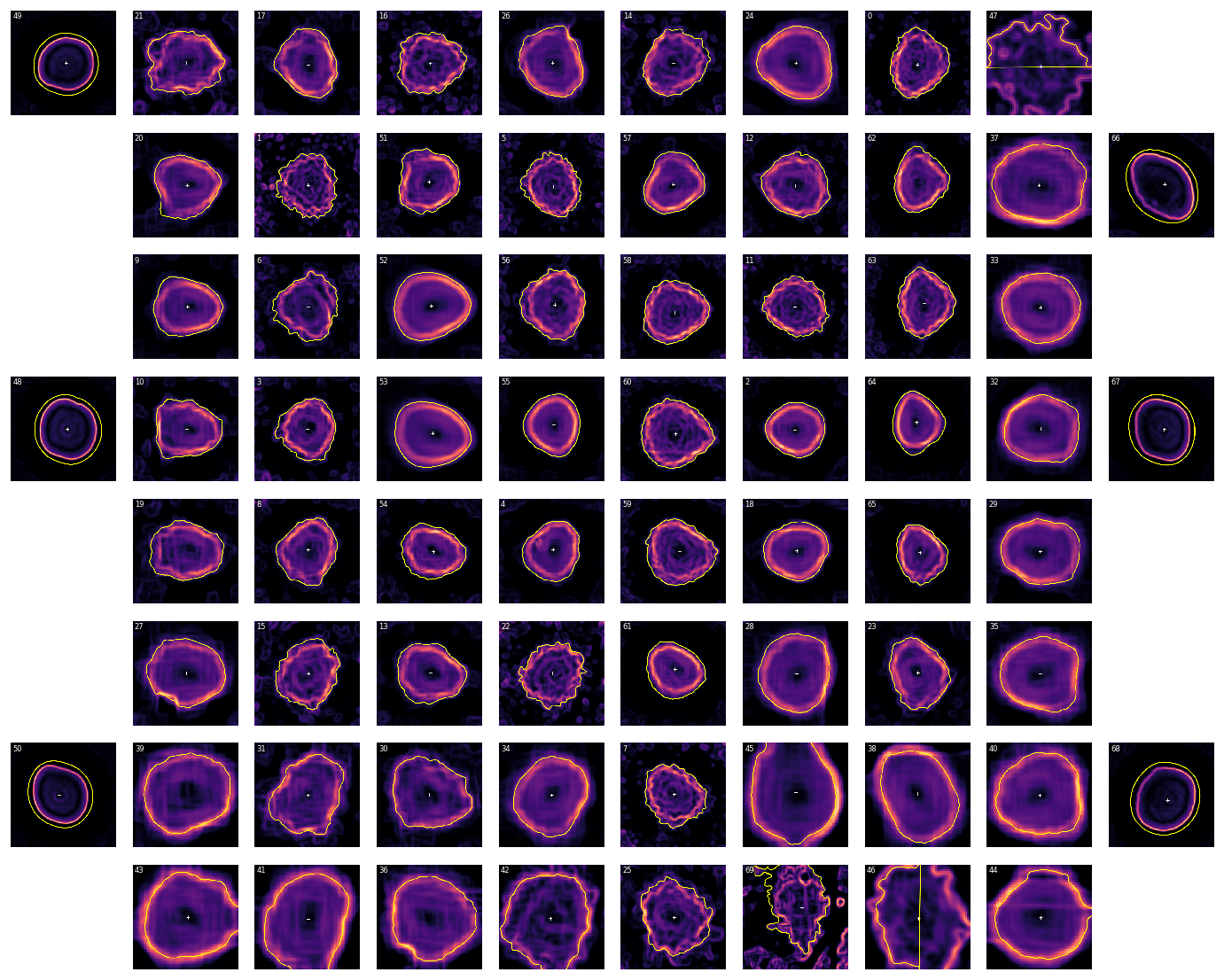}
    \caption{Human-in-the-loop agentic-AI contour co-analysis of a NIS2 image dataset from NIF. The 70 segmented neutron-image features are arranged according to their measured spatial coordinates, with each panel showing the automatically identified neutron-emission boundary and competing contour hypotheses. As in the GXD analysis, the workflow combines image conditioning, multi-scale feature extraction, autonomous candidate generation and ranking, and human supervision of ambiguous cases. The key difference is the physical target: GXD analysis emphasizes recovery of weak, nonuniform dark shell/ring structures, whereas NIS2 analysis instead identifies the bright neutron-emission envelope, using intensity-level and outward-gradient information to characterize source size, shape, and asymmetry. Thus, the same agentic-AI framework is retained while the feature evidence and contour objective are adapted to the diagnostic physics.}
    \label{fig:NISContour}
\end{figure*}

The human-in-the-loop agentic-AI workflow developed for the GXD images was
subsequently extended to NIS-2 neutron image datasets, also from NIF.  The objective of the
NIS analysis is, however, different from that of the GXD analysis.  In the GXD
images, the feature of interest was frequently a weak, spatially nonuniform
dark shell or ring embedded in a strongly varying background.  The corresponding
NIS-2 algorithm therefore searched for a finite-width dark valley and evaluated
candidate closed paths using two-sided intensity contrast, multi-scale
persistence, and geometric consistency.  In the NIS images, the dominant
structure is instead a localized bright emission distribution.  Consequently,
the same agentic framework was retained, but the image evidence and candidate
ranking were reformulated to identify the outer envelope of a bright neutron
emission feature rather than a dark shell, see Fig.~\ref{fig:NISContour}.

The NIS image was initially divided into 70 circular regions using a
pre-existing blob segmentation.  During human inspection, the original
segmentation radii were found to provide insufficient surrounding background
for some features.  An interactive preprocessing tool was therefore introduced
in which the segmentation radius could be increased while the original and
revised regions were superimposed on the full image.  A 40\% radius increase
was selected for the present analysis.  Importantly, the enlarged circles are
used only to define the image crop and do not constrain the position, radius,
or shape of the subsequently inferred neutron-emission contour. 

For each segmented image, the same family of spatial filters used in the GXD
analysis was retained.  The NIS-specific algorithm first suppresses isolated pixel-scale
fluctuations and estimates the local background and noise from the perimeter of
the segmented image.  The background is subtracted and the positive source
signal is normalized to a robust near-peak intensity.  An emission-weighted
centroid and a normalized intensity-gradient map are then calculated.  In
contrast to the GXD dark-band likelihood, for which the desired feature is
locally darker than both neighboring regions, the NIS boundary evidence rewards
a transition from a brighter interior to a darker exterior.  Strong gradients
located near the selected fractional-emission level therefore receive the
largest weight.

\subsection{AI-assisted 3D comparison of a design model and CT}

A third workflow concerns comparison of a three-dimensional (3D) design (.stl format) with a reconstructed X-ray CT volume. The broader literature has already shown that CAD priors and deep-learning reconstruction can accelerate additive-manufacturing CT characterization and improve defect analysis~\cite{Ziabari2023}. The PiMiX-specific objective is to make such operations available as provenance-tracked analysis tools that can be composed with other experimental data and, eventually, simulation or manufacturing metadata.

\begin{figure}[htbp]
    \centering
    \includegraphics[width=0.45\textwidth]{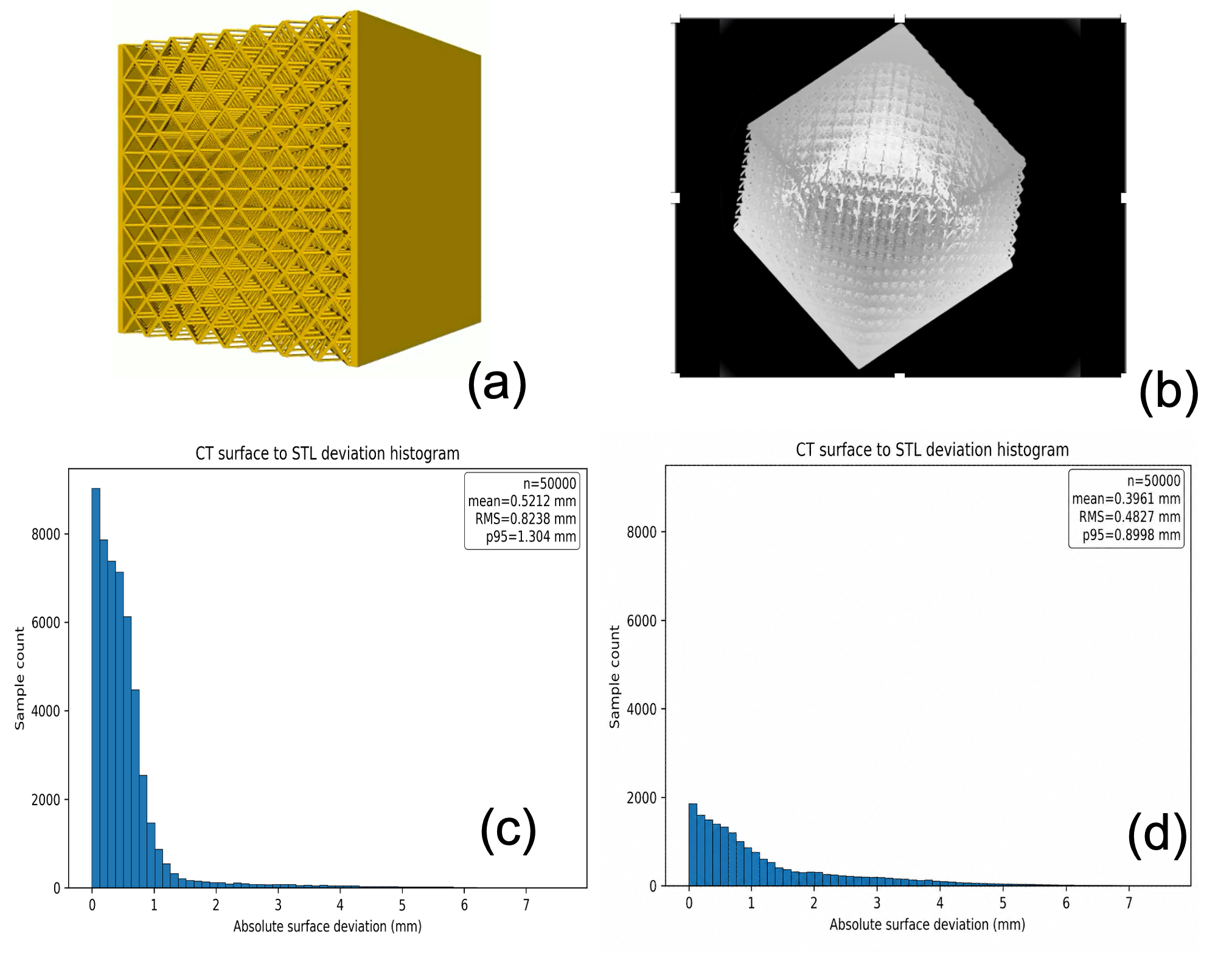}
    \caption{AI-assisted design-to-CT comparison in 3D. (a) Stereolithography (STL) design model of an additively manufactured 4340-alloy-steel lattice structure. (b) CT reconstruction of the printed structure. (c) Design/reconstruction comparison produced with a Python workflow generated or revised using an AI coding assistant. (d) Human-supervised examination of differences. Quantitative geometric-error or defect-detection metrics are not inferred from this figure alone.}
    \label{fig:CT1}
\end{figure}

Figure~\ref{fig:CT1} shows a 4340-alloy-steel lattice design, its CT reconstruction, and a prototype comparison metrics. In the workflow used for this figure, an AI coding assistant was used to generate or revise a Python library for registration/comparison, after which a human expert examined differences between the design and reconstruction and provide additional guidances for improvements. 

The workflow is currently limited by the  desktop computer memory that was used to process the design and CT data. A rigorous quantitative version of this workflow, after a memory upgrade, would follow multiple paths: specifying voxel size and CT acquisition parameters, reconstruction method, registration transform and objective function, segmentation thresholds, surface-distance or volumetric-difference metrics, defect-detection criteria, and uncertainty arising from imaging and registration. We also plan to use a calibrated reference object for the workflow validation, where many agenetic-AI-generated quantities can be compared against ground truth. The further bench-mark and code improvements aim at an important question: whether AI assistance improves throughput or accuracy relative to an established analysis pipeline.

\section{Summary and outlook\label{sec:outlook}}

PiMiX~2.0 extends the PiMiX concepts, such as MIDF, MXDF, and SXDF, to an AI-enhanced cyber-physical meta-instrument framework for
radiographic imaging and tomography (RadIT)~\cite{Wang2024:PiMiX}. The central
idea remains that measurements from different diagnostics, experiments, and
physics simulations should be treated as harmoniously coupled sources of scientific
information. The present PiMiX~2.0 framework spans three different levels
of maturity: experimentally demonstrated components, prototype integrated
workflows, and longer-term architectural targets.

At the component level, PiMiX~2.0 incorporates recent progress in multimodal
CMOS radiation imaging, simulation-assisted neutron localization, and
edge-deployed optical-neural-network inference. CMOS-based neutron imaging and
simulation-assisted learning have demonstrated sub-pixel localization
capability~\cite{LBBC:2023}. A subsequent detector-plus-localizer workflow
evaluated using both electronic GPU and optical-neural-network implementations
achieved greater than 96\% precision, greater than 98\% recall, and greater
than 97\% F1 score for neutron-event detection, together with sub-micron
localization~\cite{LZSM:2025}. These results provide quantitative examples of
AI-enhanced RadIT components with clearly defined tasks and metrics, while
full real-time integration of the CMOS detector and optical inference system
for experiments remains future work.

A major workflow-level advance reported here is the demonstration of
\emph{human-in-the-loop agentic-AI co-analysis} for X-ray and neutron images
from inertial-confinement-fusion experiments. The workflow begins with
conventional image-processing operations, including denoising, illumination
correction, spatial smoothing, Fourier-domain filtering, and iso-intensity
contouring, but extends them through iterative generation and evaluation of
competing feature hypotheses. Rather than returning a single contour obtained
from a fixed threshold or filter choice, the analysis explores alternative
preprocessing parameters and geometries, ranks candidate contours using
physics-informed evidence, estimates confidence, and presents competing
interpretations for human review and down-selection.

An important aspect of this demonstration is that the agentic architecture is
not tied to a single image morphology. For GXD X-ray data, the physical target
is a weak and spatially nonuniform dark shell or ring; the analysis therefore
uses deformable closed paths, multi-scale dark-band evidence, persistence
across preprocessing choices, and geometric consistency. For NIS-2 neutron
data, the physical objective is different, and the feature model is
reformulated to characterize a bright emission envelope using
background-subtracted fractional-emission levels, outward intensity gradients,
enclosed emission, and cross-filter persistence. The transfer of the same
analysis architecture from dark-shell inference to bright-emission inference
illustrates a key PiMiX~2.0 concept: AI can assist with the organization and
evaluation of a scientific workflow while the physical evidence used by that
workflow remains diagnostic specific.

These demonstrations represent a meaningful progression beyond carefully bounded (by human) tasks such as
AI-assisted coding, but they still need further development towards fully autonomous
 closed-loop integration. The AI assists with code generation, algorithm
modification, parameter exploration, candidate construction and ranking, and
visualization of alternatives; the scientist defines the physically meaningful
objective, evaluates difficult cases, and remains responsible for the final
scientific interpretation. 

The automated design-to-CT comparison provides a complementary workflow-level
example. In that case, an as-designed three-dimensional stereolithography model
is compared with an X-ray computed-tomography reconstruction of an additively
manufactured metal lattice. Together with the GXD and NIS demonstrations, this
example illustrates an emerging progression from AI assistance in individual
processing tasks, such as code generation or image segmentation, toward
multi-step and multi-domain scientific co-analysis in which data,
representations, algorithms, and physical criteria are coordinated within a
common framework.

The next important step is more quantitative benchmarking of these agentic
workflows and uncertainty quantification, which is currently limited in part by the computer memory used for the large datasets. The present demonstrations establish potentials, but they do not
yet establish that AI-assistance is ready for field-deployment in additive manufacturing, experiment design or operation optimization. Useful metrics include total
analysis time, number of human interventions, numerical and contour accuracy,
reproducibility, code defects, robustness to noise and preprocessing choices,
agreement with expert-selected or independently reconstructed results, and
calibration of confidence estimates. For GXD and NIS in particular, the
accuracy of image-space feature extraction should also be distinguished from
the uncertainty of any downstream physical inference. 

A longer-term objective of PiMiX~2.0 is iterative DF, in which a physics model
or digital twin generates predicted diagnostic observables. The predictions
are compared with experimental measurements, and model parameters are updated with
quantified uncertainty. The resulting state estimate informs subsequent
analysis or experimentation. Such a capability may be scientifically more
demanding than connecting a LLM to an experimental control
interface. A credible implementation requires an explicit model
parameterization, calibrated measurement uncertainties, a stated likelihood or
discrepancy metric, treatment of model inadequacy, and well-defined constraints
on which parameters or experimental variables may be modified.

The agentic workflows demonstrated here also identify several near-term
engineering requirements. PiMiX~2.0 is presently a heterogeneous collection of
demonstrated components, prototype workflows, and architectural targets rather
than a single deployed software package. Reproducible agentic analysis will
therefore require machine-readable data and metadata contracts, registered and
versioned analysis tools, explicit detector geometry, provenance records,
automated tests, and well-defined interfaces between experimental data,
simulation codes, and learned models. The GXD and NIS workflows provide useful
prototypes for such an architecture because they already couple preprocessing,
feature construction, candidate generation, quantitative ranking, confidence
estimation, visualization, and human override.

PRISM (PiMiX Radiographic Intelligence and Scientific Model) represents a
longer-term `AI Master" (AIM) within this architecture. It is envisioned as
a RadIT scientific foundation model capable of relating images, spectra,
geometry, detector response, simulations, metadata, and scientific text.
PRISM is not evaluated in the present work as a completed foundation model.
Substantial research remains necessary in training-data governance,
cross-diagnostic and cross-facility generalization, representation of geometry
and uncertainty, synthetic-to-experimental domain shift, interpretability,
quantitative benchmarking, and robustness to out-of-distribution conditions.
Existing scientific and imaging foundation models provide precedent for this
direction~\cite{Pai2024,Niu2025}, but do not by themselves establish that a
RadIT-specific model will generalize across instruments and facilities.

Autonomous experimental control is correspondingly a roadmap endpoint rather
than a capability claimed here. A subsequent hardware-in-the-loop study could
evaluate whether inferred states or agent-generated recommendations can inform
detector settings, timing, target positioning, or other bounded experimental
choices while retaining human authorization. High-energy-density and fusion
facilities introduce stringent requirements for safety, cybersecurity,
latency, reliability, operational interlocks, and fail-safe behavior. Prior
work in autonomous laboratories demonstrates that tool orchestration and
closed-loop experimentation are possible in other scientific
domains~\cite{Szymanski2023ALab,Boiko2023Coscientist}, but the appropriate path
for PiMiX is staged: AI assistance, human-supervised agentic co-analysis,
offline recommendation, simulation and hardware-in-the-loop validation,
advisory operation, and only then carefully bounded automation.

\begin{acknowledgments}
This work is supported in part by the ICF program (managers: Ann Satsangi and Joseph Smidt) of the Office of Experimental Sciences and the Los Alamos National Laboratory LDRD program. The work led by The University of Texas at Austin is supported in part by the AFOSR MURI research center on Energy-efficient Optical Interconnects and Computing.
\end{acknowledgments}

\end{document}